\documentclass[twoside,fleqn]{article}
\usepackage{espcrc2}
\usepackage{amssymb}
\usepackage[dvips]{epsfig}
\newcommand{\be}{\begin{equation}}
\newcommand{\ee}{\end{equation}}
\newcommand{\bea}{\begin{eqnarray}}
\newcommand{\eea}{\end{eqnarray}}
\newcommand{\bean}{\begin{eqnarray*}}
\newcommand{\eean}{\end{eqnarray*}}
\newcounter{saveeqn}
\newcommand{\aeqn}{\stepcounter{equation}\setcounter{saveeqn}
                  {\value{equation}}%
\setcounter{equation}{0}%
\renewcommand{\theequation}
             {\mbox{\arabic{saveeqn}\alph{equation}}}}
\newcommand{\reqn}{\setcounter{equation}{\value{saveeqn}}%
\renewcommand{\theequation}{\arabic{equation}}}
\newcounter{doeqn}

\newcommand{\T}{\textstyle}
\newcommand{\dts}[2]{#1_{{\scriptsize \mbox{#2}}}}

\newcommand{\mvl}[1]{\left\langle#1\right\rangle}

\newcommand{\Smu}{\tilde{\partial}_{\mu}}
\newcommand{\Snu}{\tilde{\partial}_{\nu}}
\newcommand{\Sz}{\tilde{\partial}_{0}}

\newcommand{\gfv}{\gamma_5}

\newcommand{\fa}{f_{{\scriptsize \mbox{A}}}}
\newcommand{\fai}{f_{{\scriptsize \mbox{A}}}^{{\scriptsize \mbox{I}}}}
\newcommand{\far}{f_{{\scriptsize \mbox{A}}}^{{\scriptsize \mbox{R}}}}
\newcommand{\fp}{f_{{\scriptsize \mbox{P}}}}
\newcommand{\kv}{k_{{\scriptsize \mbox{V}}}}
\newcommand{\kvi}{k_{{\scriptsize \mbox{V}}}^{{\scriptsize \mbox{I}}}}
\newcommand{\kvr}{k_{{\scriptsize \mbox{V}}}^{{\scriptsize \mbox{R}}}}
\newcommand{\kt}{k_{{\scriptsize \mbox{T}}}}
\newcommand{\csw}{c_{{\scriptsize \mbox{sw}}}}
\newcommand{\ca}{c_{{\scriptsize \mbox{A}}}}
\newcommand{\cv}{c_{{\scriptsize \mbox{V}}}}
\newcommand{\cX}{c_{{\scriptsize \mbox{X}}}}
\newcommand{\ct}{c_{{\scriptsize \mbox{t}}}}
\newcommand{\ctt}{\tilde{c}_{{\scriptsize \mbox{t}}}}
\newcommand{\ba}{b_{{\scriptsize \mbox{A}}}}
\newcommand{\bV}{b_{{\scriptsize \mbox{V}}}}
\newcommand{\bP}{b_{{\scriptsize \mbox{P}}}}

\newcommand{\bX}{b_{{\scriptsize \mbox{X}}}}
\newcommand{\za}{Z_{{\scriptsize \mbox{A}}}}
\newcommand{\zp}{Z_{{\scriptsize \mbox{P}}}}
\newcommand{\zv}{Z_{{\scriptsize \mbox{V}}}}

\newcommand{\zX}{Z_{{\scriptsize \mbox{X}}}}
\newcommand{\mq}{m_{{\scriptsize \mbox{q}}}}

\newcommand{\mbar}{\overline{m}}
\newcommand{\mpcac}{\dts{m}{PCAC}}
\newcommand{\mps}{\dts{m}{PS}}
\newcommand{\fps}{\dts{f}{PS}}

\newcommand{\mpi}{m_{\pi}}
\newcommand{\fpi}{f_{\pi}}

\newcommand{\mV}{\dts{m}{V}}
\newcommand{\fV}{\dts{f}{V}}
\newcommand{\mrh}{m_{\rho}}
\newcommand{\frh}{f_{\rho}}
\newcommand{\kapc}{\kappa_{{\scriptsize \mbox{c}}}}
\newcommand{\amu}{A_{\mu}^a}
\newcommand{\az}{A_0^a}

\newcommand{\amui}{(A_{{\scriptsize \mbox{I}}})_{\mu}^a}

\newcommand{\amur}{(A_{{\scriptsize \mbox{R}}})_{\mu}^a}

\newcommand{\p}{P^a}
\newcommand{\pr}{P^a_{{\scriptsize \mbox{R}}}}
\newcommand{\tmunu}{T^a_{\mu\nu}}
\newcommand{\tkz}{T^a_{k0}}
\newcommand{\vmu}{V_{\mu}^a}

\newcommand{\vmui}{(V_{{\scriptsize \mbox{I}}})_{\mu}^a}

\newcommand{\vmur}{(V_{{\scriptsize \mbox{R}}})_{\mu}^a}

\newcommand{\bu}{{\mbox{\textbf{u}}}}
\newcommand{\bv}{{\mbox{\textbf{v}}}}

\newcommand{\bus}{{\scriptsize \mbox{\textbf{u}}}}
\newcommand{\bvs}{{\scriptsize \mbox{\textbf{v}}}}

\newcommand{\obi}{\mathcal{O}^a}
\newcommand{\obf}{\mathcal{O}'^a}
\newcommand{\qbi}{\mathcal{Q}^a_k}

\newcommand{\zeb}{\bar{\zeta}}
\newcommand{\Or}{\mbox{O}}
\title{
{
\vspace{-3.0cm} \normalsize \hfill
\parbox{30mm}{DESY 98-123\\hep-lat/9809002}
}\\[15mm]
Scaling tests in $\Or(a)$--improved quenched lattice QCD%
\thanks{Talk given at the International Symposium on Lattice Field
Theory, July 13--18, 1998, in Boulder, USA.}
}
\author{
Jochen Heitger\address{
Deutsches Elektronen-Synchrotron DESY Zeuthen, Platanenallee~6,
D-15738 Zeuthen, Germany}
}
\begin{document}
%
%%%%%    slight change in table style, J.H. 1996)    %%%%
\makeatletter
\long\def\@maketablecaption#1#2{\vskip 10mm #1. #2\par}
\makeatother
%%%%%%%%%%%%%%%%%%%%%%%%%%%%%%%%%%%%%%%%%%%%%%%%%%%%%%%%%
%
\begin{abstract}
We present a scaling investigation of renormalized correlation
functions in $\Or(a)$--improved quenched lattice QCD.
As one observable the renormalized PCAC quark mass is considered,
others are constructed such that they become the vector meson mass,
and the pseudoscalar and vector meson decay constants in large volume.
Presently, we remain in intermediate volume,
$(0.75^3\times1.5)\,\,\mbox{fm}^4$, and study the approach to the
continuum limit.
\end{abstract}
\maketitle
%
%%%%%%%%%%%%%%%%%%%%%%%%%%%%%%%%%%%%%%%%%%%%%%%%%%%%%%%%%%%%%%%%%%%%%%%%%
%
\section{Introduction}
Discretization errors of $\Or(a)$ in lattice QCD can be removed via
a systematic approach based on the Symanzik improvement
programme \cite{S83}, which adds appropriate higher-dimensional
operators to action and fields of interest \cite{LSSW96S98}.
Exploiting chiral symmetry restoration and certain current Ward
identities on the lattice, a (mostly) non-perturbative
$\Or(a)$--improvement for action and quark currents as well as their
renormalization has been achieved in the quenched case within this
framework \cite{LSSWW97LSSW97,GS98,DP98}.

Thus one is not only interested in the influence of improvement on
agreement of lattice data with experiment at fixed $\beta$--values,
but also in the quality of scaling and the size of its violation.
In this context it was reported \cite{GHPRSSS98W98} that at
$a\simeq 0.1\,\,\mbox{fm}$ the residual $\Or(a^2)$ lattice artifacts
may be fairly large e.g.~for $f_Kr_0$ ($\sim$ 10 \%), while they are
already very small for other quantities like
$\mrh/\sqrt{\sigma}$ ($\sim$ 2 \%).
Restricting to an intermediate volume, we therefore examined the
impact of $\Or(a)$--improvement thoroughly and with high accuracy for
different observables.
\section{Fermionic correlation functions}
Consider correlation functions in the Schr\"odinger functional (SF)
\cite{LNWW92S94} with all details found in
\cite{LSSW96S98,LSSWW97LSSW97}.
We use $\zeta(\zeb)$ as $x_0=0$ boundary (anti-)quark fields in
$\obi=a^6\sum_{\bus,\bvs}\zeb(\bu)\gfv\frac{\tau^a}{2}\zeta(\bv)$ and
$\qbi=a^6\sum_{\bus,\bvs}\zeb(\bu)\gamma_k\frac{\tau^a}{2}\zeta(\bv)$,
and axial (vector) current $\amu(\vmu)$ and pseudoscalar (tensor)
density $\p(\tmunu)$ to form the expectation values
\aeqn\bea
\fa(x_0)
& = &-\frac{1}{3}\,\mvl{\az(x)\,\obi}
\label{cor_fa}\\
\fp(x_0)
& = &-\frac{1}{3}\,\mvl{\p(x)\,\obi}
\label{cor_fp}\\
f_1
& = &-\frac{1}{3L^6}\,\mvl{\obf\obi}
\label{cor_f1}\\
\kv(x_0)
& = &-\frac{1}{9}\,\mvl{V_k^a(x)\,\qbi}
\label{cor_kv}\\
\kt(x_0)
& = &-\frac{1}{9}\,\mvl{\tkz(x)\,\qbi}\,.
\label{kor_kt}
\eea\reqn
Given the improvement coefficients $\ca$ and $\cv$, the improved currents
$\amui\equiv\amu+a\ca\Smu\p$ and $\vmui\equiv\vmu+a\cv\Snu\tmunu$
lead to define the corresponding fermionic correlation functions
\aeqn\bea
\fai(x_0)
& = &\fa(x_0)+a\ca\Sz\fp(x_0)
\label{cor_fa_i}\\
\kvi(x_0)
& = &\kv(x_0)+a\cv\Sz\kt(x_0)\,,
\label{cor_kv_i}
\eea\reqn
where the symmetrized lattice derivative acts as usual as
$\Smu f(x)=[f(x+a\hat{\mu})-f(x-a\hat{\mu})]/2a$.
\section{Observables under study and results}
Employing a mass-independent renormalization respecting
$\Or(a)$--improvement, the quantities
\aeqn\bea
\amur
& = &\za(1+\ba a\mq)\amui
\label{cur_axl_r}\\
\vmur
& = &\zv(1+\bV a\mq)\vmui
\label{cur_vec_r}\\
\pr
& = &\zp(1+\bP a\mq)\p
\label{den_psc_r}
\eea\reqn
induce the renormalized correlation functions
\aeqn\bea
\far(x_0)
& = &\za(1+\ba a\mq)\fai(x_0)
\label{cor_fa_r}\\
\kvr(x_0)
& = &\zv(1+\bV a\mq)\kvi(x_0)
\label{cor_kv_r}
\eea\reqn
with $\zX$ and $\bX$, X=A,V,P, which are \emph{not} functions of the
quark mass $a\mq=(1/\kappa-1/\kapc)/2$.
Now we construct the following set of observables:
\bea
m(x_0)
& = &\frac{\Sz\fa(x_0)+a\ca\Sz^2\fp(x_0)}{2\fp(x_0)}
\label{mcurr_x0}\\
\mps(x_0)
& = &\frac{\Sz\fp(x_0)}{\fp(x_0)}\,,\quad
     \mpi=\mps({\T \frac{T}{2}})
\label{mpi_x0}\\
\mV(x_0)
& = &\frac{\Sz\kvi(x_0)}{\kvi(x_0)}\,,\quad
     \mrh=\mV({\T \frac{T}{2}})
\label{mrho_x0}\\
a\fps(x_0)
& \propto &\frac{\far(x_0)}{\sqrt{f_1}}\,,\quad
           \fpi=\fps({\T \frac{T}{2}})
\label{fpi_x0}\\
\fV^{-1}(x_0)
& \propto &\frac{\kvr(x_0)}{\sqrt{f_1}}\,,\quad
           \frh^{-1}=\fV^{-1}({\T \frac{T}{2}})\,.
\label{frho_x0}
\eea
The division by $\sqrt{f_1}$ cancels the renormalizations of the
boundary quark fields ensuring that $\fpi$,$\frh$ are scaling
quantities, and the proportionality constants in
eqs.~(\ref{fpi_x0}),(\ref{frho_x0}) are such that these ratios
turn, as $T\rightarrow\infty$, into the familiar matrix elements, which
define the $\pi$ and $\rho$ meson decay constants.
The renormalized PCAC (current) quark mass in the SF scheme is obtained
as \cite{CGLSSWW98}
\be
\mbar=\frac{\za}{\zp}\,\mpcac\,,\quad
\mpcac=m(x_0)\,\Big|_{x_0=\frac{T}{2}}\,,
\label{mpcac}
\ee
where $(\ba-\bP)a\mq$ may be neglected \cite{DP98}.

For the analysis we use $\csw$, $\cX$ and $\zX$ non-perturbatively
determined in \cite{LSSWW97LSSW97,GS98,W99} for $\beta\ge6.0$, while
$\ba$, $\bV$ and $\bP$, as well as the SF specific
improvement coefficients of the boundary counterterms $\ct$ and $\ctt$,
are taken from 1--loop perturbation theory 
\cite{LSSW96S98,LW96SW97,LNWW92S94}.  

%
%%% Beginn Tabelle (einspaltig) %%%
\begin{table}[htb]
\begin{center}
\vspace{-0.5cm}
\begin{tabular*}{7.5cm}{ccccc}
\hline
  $L/a$ & $\beta$ & $\kappa$ & $L/r_0$  & $\mpi L$  \\
\hline
  8     & 6.0     & 0.13458  & 1.490(6) & 2.004(9)  \\
  10    & 6.14    & 0.13538  & 1.486(7) & 1.946(14) \\
  12    & 6.26    & 0.13546  & 1.495(7) & 2.050(16) \\
  16    & 6.48    & 0.13541  & 1.468(8) & 1.991(15) \\
\hline
\end{tabular*}
\parbox{7.5cm}{
\vspace{-0.7cm}
\caption{\label{ParTab} \sl Simulation points for the LCP studied.}
}
\vspace{-1.2cm}
\end{center}
\end{table}
%%% Ende Tabelle %%%
%
The strategy was then to keep a finite physical volume and the quark
mass fixed by prescribing the geometry $T/L=2$, `pion' mass $\mpi L=2.0$
and spatial lattice size $L/r_0=1.49$ using the recent results on the
hadronic scale $r_0/a$ \cite{GSW98}, see table~\ref{ParTab}.
%
%%% Beginn Figur %%%
\begin{figure}[htb]
\begin{center}
\epsfig{file=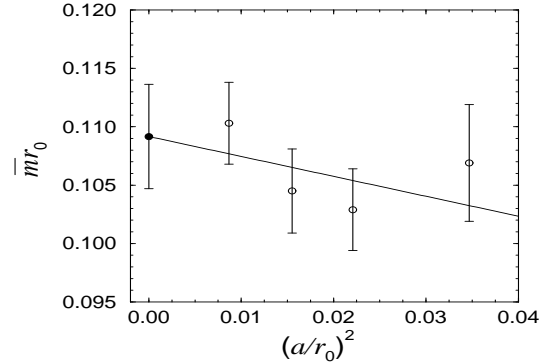,
        width=7.0cm,height=5.0cm}
\parbox{7.5cm}{
\vspace{-0.9cm}
\caption{\label{MbarPlot} \sl Scaling of the PCAC quark mass
                              in intermediate volume with SF
                              boundary conditions.}
}
\vspace{-1.25cm}
\end{center}
\end{figure}
%%% Ende Figur %%%
%
%
%%% Beginn Figur %%%
\begin{figure}[htb]
\begin{center}
\epsfig{file=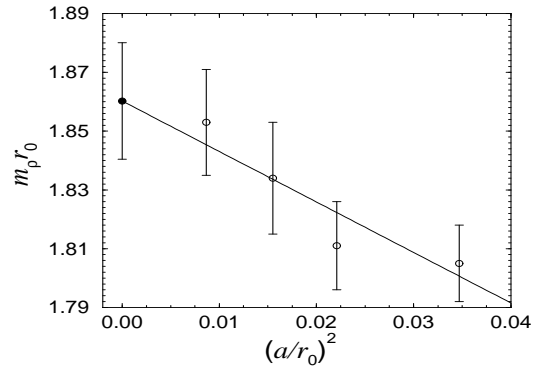,
        width=7.0cm,height=5.0cm}
\parbox{7.5cm}{
\vspace{-0.9cm}
\caption{\label{MrhoPlot} \sl Same as figure~\ref{MbarPlot} but for
                              the rho mass.}
}
\vspace{-1.5cm}
\end{center}
\end{figure}
%%% Ende Figur %%%
%
%
%%% Beginn Figur %%%
\begin{figure}[htb]
\begin{center}
\epsfig{file=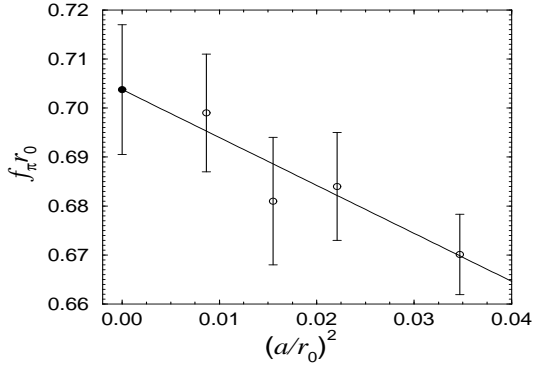,
        width=7.0cm,height=5.0cm}
\parbox{7.5cm}{
\vspace{-0.9cm}
\caption{\label{FpiPlot} \sl As figure~\ref{MbarPlot} but for
                             the $\pi$ decay constant.}
}
\vspace{-1.5cm}
\end{center}
\end{figure}
%%% Ende Figur %%%
%
%
%%% Beginn Figur %%%
\begin{figure}[htb]
\begin{center}
\epsfig{file=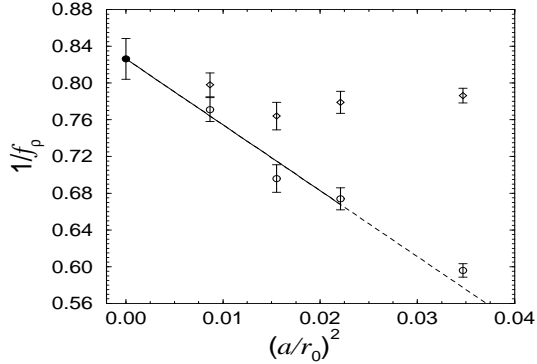,
        width=7.0cm,height=5.0cm}
\parbox{7.5cm}{
\vspace{-0.9cm}
\caption{\label{FrhoPlot} \sl As figure~\ref{MbarPlot} but for
                              the $\rho$ decay constant.
                              Diamonds refer to an analysis
                              with $\cv=0$.}
}
\vspace{-1.25cm}
\end{center}
\end{figure}
%%% Ende Figur %%%
%
This yields an intermediate volume of $(0.75^3\times1.5)\,\,\mbox{fm}^4$,
moving on a line of constant physics (LCP) in parameter space with lattice
resolutions ranging from 0.1 to 0.05 fm.
As an important prerequisite for the reliability of the scaling test we
verified by variation of the 1--loop values as simulation input that
the dependence on $\ct,\ctt$ is small enough to be neglected.
The leading scaling violations should therefore be $\Or(a^2)$.
Any small mismatch with the renormalization conditions on $\mpi L$ and
$L/r_0$ was corrected by an estimation of the corresponding slopes.
Finally, we performed extrapolations to the continuum limit, assuming
convergence with a rate $\propto a^2$.
%
%%% Beginn Tabelle (einspaltig) %%%
\begin{table}[htb]
\begin{center}
\vspace{-0.75cm}
\begin{tabular*}{7.5cm}{ccccc}
\hline
  $\mbar r_0$ & $\mrh r_0$ & $\fpi r_0$ & $1/\frh$  \\
\hline
  0.1092(45)  & 1.860(20)  & 0.704(13)  & 0.826(22) \\
  2.1 \%      & 3.0 \%     & 4.8 \%     & 28 \%     \\
\hline
\end{tabular*}
\parbox{7.5cm}{
\vspace{-0.7cm}
\caption{\label{CLimTab} \sl Continuum limits and their percentage
                             deviations from $\beta=6.0$ 
                             ($a\simeq 0.1\,\,\mbox{fm}$).}
}
\end{center}
\end{table}
%%% Ende Tabelle %%%
%

The fits are displayed in figures~\ref{MbarPlot} -- \ref{FrhoPlot},
where the total error is always dominated by the uncertainties of the
renormalization factors $\zX$.
One observes the leading corrections to the continuum to be compatible
with $\Or(a^2)$.
Moreover, it can be inferred from table~\ref{CLimTab} that the difference
of the continuum limits from the values at $\beta=6.0$
($a\simeq 0.1\,\,\mbox{fm}$) is below 5 \% in the improved theory.
The only exception is the inverse `rho' meson decay constant, whose
slope is quite large; for that reason we discard the $\beta=6.0$ point
to extrapolate to the continuum limit.
\section{Discussion and outlook}
Our numerical simulations of renormalized and improved correlation
functions show an overall behaviour completely consistent with being
linear in $a^2$ at $\beta\ge6.0$ for all quantities under consideration.
Changing $a$ by a factor 2 gives very stable continuum extrapolations.
 
But the residual $\Or(a^2)$--effect in $1/\frh$ is still large at
$a\simeq 0.1\,\,\mbox{fm}$ ($\lesssim$ 30 \%).
Here we find that artificially setting $\cv=0$ results in data with an
overall weaker dependence on $a$.
However, as indicated in figure~\ref{FrhoPlot}, for such a $\cv$--value
the functional form of the $a$--effects seems no longer compatible
with $a^2$.
Additionally, a chiral Ward identity is badly violated for $\cv=0$ at 
$\Or(a)$--level \cite{GS98}.
Thus there is no choice of $\cv$, which makes $\Or(a^2)$--effects small
in both channels.
This example clearly illustrates that even in the $\Or(a)$--improved
theory the remaining $\Or(a^2)$--effects have to be assessed by varying
the lattice spacing.

This work is part of the ALPHA collaboration research programme.
We thank DESY for allocating computer time to this project.
%
%%%%%%%%%%%%%%%%%%%%%%%%%%%%%%%%%%%%%%%%%%%%%%%%%%%%%%%%%%%%%%%%%%%%%%%%%
%
% bibliography
%

%
\end{document}